# RFID Technology Based Attendance Management System

Sumita Nainan[1], Romin Parekh[2], Tanvi Shah[3]

[1] Department of Electronics & Telecommunication Engineering, NMIMS University
Mumbai, Maharashtra 400 056. INDIA

[2] Department of Computer Engineering, NMIMS University
Mumbai, Maharashtra 400 056. INDIA

[3] Department of Computer Engineering, NMIMS University
Mumbai, Maharashtra 400 056. INDIA

**Abstract**
RFID is a nascent technology, deeply rooted by its early developments in using radar[1] as a harbinger of adversary planes during World War II. A plethora of industries have leveraged the benefits of RFID technology for enhancements in sectors like military, sports, security, airline, animal farms, healthcare and other areas. Industry specific key applications of this technology include vehicle tracking, automated inventory management, animal monitoring, secure store checkouts, supply chain management, automatic payment, sport timing technologies, etc. This paper introduces the distinctive components of RFID technology and focuses on its core competencies: scalability and security. It will be then supplemented by a detailed synopsis of an investigation conducted to test the feasibility and practicality of RFID technology.

## 1. Introduction

RFID, which stands for Radio Frequency Identification, is an automatic identification technology used for retrieving from or storing data on to RFID Tags without any physical contact [1]. An RFID system primarily comprises of RFID Tags, RFID Reader, Middleware and a Backend database. RFID Tags are uniquely and universally identified by an identification sequence, governed by the rubrics of EPCglobal Tag Data Standard[2]. A tag can either be passively activated by an RFID reader or it can actively transmit RF signals to the reader [3]. The RFID reader, through its antenna, reads the information stored on these tags when it's in its vicinity. The reader, whose effective range is based on its operational frequency, is designed to operate at a certain frequency. The operational frequency of the reader ranges from 125 KHz – 2.4 GHz [5]. The Middleware encompasses all those components that are responsible for the transmission of germane information from the reader to the backend management systems [8]. The Middleware can include hardware components like cables and connectivity ports and software components like filters that monitor network performance of the system [2, 9]. The Backend database stores individual tag identifiers to uniquely identify the roles of each tag. The database stores record entries pertaining to individual tags and its role in the system application. The RFID system is interdependent on its core components to achieve maximum efficiency and optimum performance of the application. Due to its high degree of flexibility, the system can be easily adopted for an array of applications ranging from small scale inventory cabinets to multifarious and highly agile supply chain management systems [4, 6]. Although, the cost of incorporating this technology has restricted its outreach, the technology promises to have untapped potential [10, 11].

## 2. Evolution of RFID

The success of RFID technology primarily centres on the advent of radio technology [12]. The developments in radio technology were a prerequisite to harness the essence of RFID technology. There is significant growth over the past couple of decades in this technology (see figure 1). RFID technology is rife in modern industries that demand data integrity and high efficiency of the system. This technology is used for tracking vehicles and goods, courier services and luggage handling [18]. Other applications include animal tracking, secure toll payments, inventory management systems, access control mechanisms, etc. Figure 1 depicts the evolution of RFID technology.

---

[1] Radio Detection and Ranging is a communication medium to subliminally detect objects that are miles away, invisible to the naked eye.
[2] It defines the guidelines of how key identifiers must be encoded on the tag to define industry based standardization.

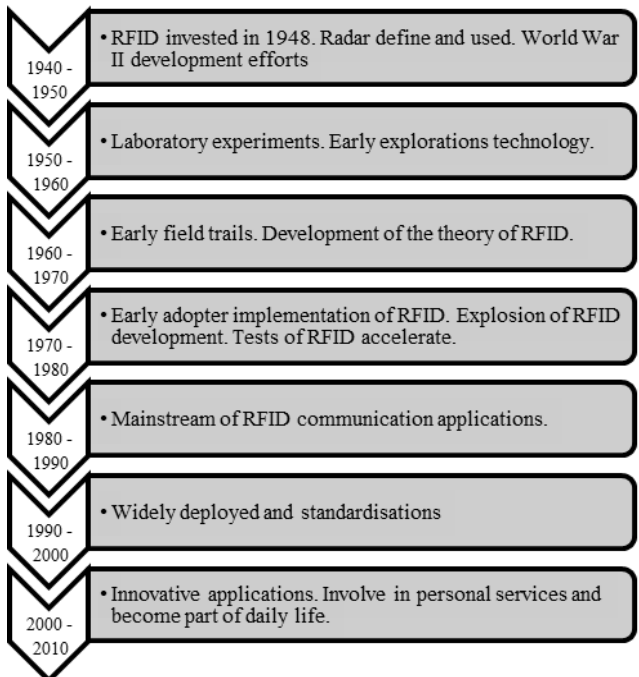

Fig. 1 A depiction of the evolution of RFID technology adapted from [22].

## 3. Components of an RFID System

An RFID system consists of various components that are connected to one another by a dedicated communication path (see figure 2). The individual components are integrated into the system to implement the benefits of RFID solution [15]. The list of components is as follows:

- Tags – an object that is attached to any product and uses a unique sequence of characters to define it. It comprises of a chip and the antenna.
- Antenna – it is responsible for the transmission of information between the reader and tag using radio waves.
- Reader – a scanning device that uses the antenna to realise the tags that are in its vicinity. It transmits signals at a certain frequencies.
- Middleware – it is a communication interface to interpret and process data being fed by the readers into information. It takes into account all relevant ports of communication and a software application to represent this information.
- Backend database – a repository of information, which is designed specific to the application. The database stores records of data specific to individual tags.

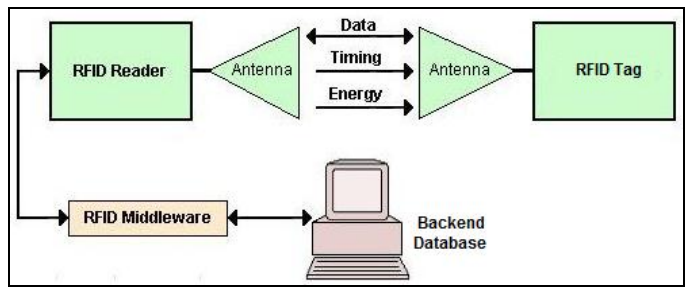

Fig. 2 Components of an RFID system

### 2.1 Tags

A tag consists of a microchip that stores a unique sequence identifier that is useful in identifying objects individually. The sequence is a numeric serial, which is stored in the RFID memory. The microchip includes minute circuitry and an embedded silicon chip [14, 18]. The tag memory can be permanent or re-writable, which can be re-programmed electronically by the reader multiple times. Tags are designed specific to its applications and environment. For example, paper-thin tags are attached to books in a library management system [12].

Tags are available in various shapes and sizes (see figure 3). Tags that are initiated by the reader are known as Passive tags, whilst those that do not require external initiation are called Active tags. A Semi-Passive tag exists, which has the features of both Active and Passive tags [21]. Each tag type has its distinct characteristics, which are discussed in table 1.

Tags are operable on Microwave (2.4 – 2.5 GHz), Ultra High Frequency (UHF) (860 – 1500 MHz), High Frequency (HF) (13.56 MHz) and Low Frequency (LF) (125 kHz) [22].

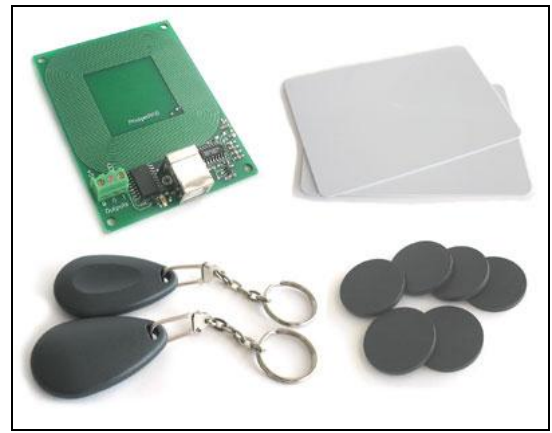

Fig. 3 Types of RFID Tags

Table 1: Features of Types of Tags

| Feature | Type of Tag | | |
|---|---|---|---|
| | **Passive** | **Active** | **Semi – Passive** |
| Read Range | Short (up to 10m) | Long (up to 100m) | Long (up to 100m) |
| Battery | No | Yes | Yes |
| Lifespan | Up to 20 years | Between 5-10 years | Up to 10 years |
| Cost | Cheap | Very expensive | Expensive |
| Availability | Only in field of reader | Continuous | Only in field of reader |
| Storage | 128 bytes read/ write | 128 Kbytes read/ write | 128 Kbytes read/ write |
| Application | EZ-Pass toll payment booths | Monitor the condition of fresh produce | Measurement of temperature periodically |

### 2.2 Antenna

The antenna is medium through which the tag and reader communicate with each other. It antenna can activate a passive tag and transfer data by emitting wireless impulses that has electromagnetic properties [20]. The antenna comes in various designs (see figure 4). They come in following types: (1) Stick antennas, (2) Di-pole or multi-pole antennas, (3) Beam-forming or phased-array element antennas, (4) Circular polarized, (5) Gate antennas, (6) Patch antennas, (7) Linear polarized, (8) Adaptive antennas, and (9) Omni directional antennas [19].

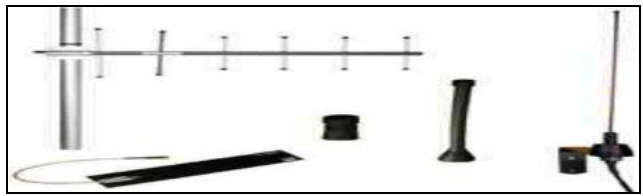

Fig. 4 Types of antenna [7]

### 2.3 Reader

The reader is the most fundamental part of the RFID system. It reads raw data from the tag and transmits it to the Middleware for further processing [16]. The reader attempts to interrogate the tags at varying frequencies. The reader communicates by transmitting a beam of impulses, which encapsulate commands to the tag and listens for the tag's response [14]. The reader also contains built in anti-collision processes, which allows the reader to read multiple tags simultaneously [15]. The reader is connected to the computer for data processing via a USB cable or over a wireless connection.

### 2.4 Middleware

The middleware is an interface required to manage the flow of data from the reader and to transmit it efficiently to the backend database management systems [18]. The middleware monitors the number of tags present in the system and extracts relevant information from the readers [12]

### 2.5 Backend Database

The backend database primarily deals with the storage of relevant information recorded by the reader and communicated by the middleware [16]. For example, the middleware in an automated security control system will store all tag readings taken by the reader in the database. This helps create log entries for the system [19].

## 4. Research

RFID technology has a widened horizon as it transcends into an era of emerging applications [1]. A detailed research must be conducted to assay the limitations and feasibility of implementing an RFID system [3, 4]. This paper focuses on the development of an attendance management system using RFID technology to monitor the attendance for a group of students [2]. This paper attempts to evaluate the benefits of implementing RFID technology to an existing system. The implementation of RFID in student management will provide additional capabilities like high efficiency and overall ease in management of the system [11]. The objectives of the research should be clearly organised to successfully develop the system.

## 5. Application Description

The primary aim of the research is to uniquely identify individual students based on their unique tag identifiers [22]. The research should shower light on how scalable and efficient the system is [15]. A systematic and serialised approach is required to solve this conundrum. The key characteristics of the application include:
- Perform automated attendance
- Generate report of attendees for a particular course
- Error free tag identifier detection
- Easy scalability to incorporate more records
- Integrity and security in data storage

This paper concentrates on the principal purpose to overcome the human errors while recording student attendance and the creation of a data centric student attendance database system with an improved overall efficiency. The application graphical user interface (GUI) is designed using Visual Basic 6.0[3] and Microsoft Access is used as the database provider. The Atmel[4] AT89S52 is the heart of the system, which is a low-power high performance CMOS 8-bit microcomputer with 8K bytes of downloadable flash programmable and erasable read only memory [11]. It is operable in two modes namely (1) Idle mode and (2) Power down mode [9, 11]. The microcontroller can be programmed with the 80C51 instruction set along with additional standardised features like:
- 256 bytes of RAM[5]
- 32 Input/ Output data lines
- Three 16bit timers/ counters
- SPI[6] serial interface
- Power off flag

The circuit contains a 16x2 LCD[7] display panel, which is the output device of the system [17, 19]. It displays the user's information when the stored tag is read by the reader. The serial interface allows connectivity to a local database for data storage and retrieval [20]. The input to the system is the unique tag identifier stored in the RF tag, which is sensed by the reader [21]. The components are mounted on the printed circuit board for interconnectivity between them.

---

[3] Visual Basic is a high level programming language developed by Microsoft.
[4] Atmel Corporation is a worldwide leader in the design and manufacture of microcontrollers, capacitive touch solutions, advanced logic, mixed-signal, non-volatile memory and radio frequency (RF) components [9].
[5] RAM is an acronym for Random Access Memory, which is a volatile type of memory required by the computer at runtime.
[6] SPI is an acronym for Serial Peripheral Interface, which is 4-wire serial communications interface to provide stable rate of data transferring [6].
[7] LCD stands for Liquid Crystal Display, which paints a picture on the screen by correcting the orientation of the liquid crystals by applying alternating currents [9].

The software module of the middleware processes the raw data fed in by the hardware circuit. The raw data fed into the middleware are:
- Unique tag sequence number
- Timestamp of data entry

The middleware obtains the unique identifier from the reader and compares it with the list of stored tags. If the identifier sequence is present, then the details are fetched and displayed on the LCD display and the GUI (see figure 5). If the identifier is not present then a new record is created with the corresponding timestamp and it is stored in the database. The student will be prompted to fill in the following details:
- Name
- Course details
    → Course
    → Stream
    → Trimester

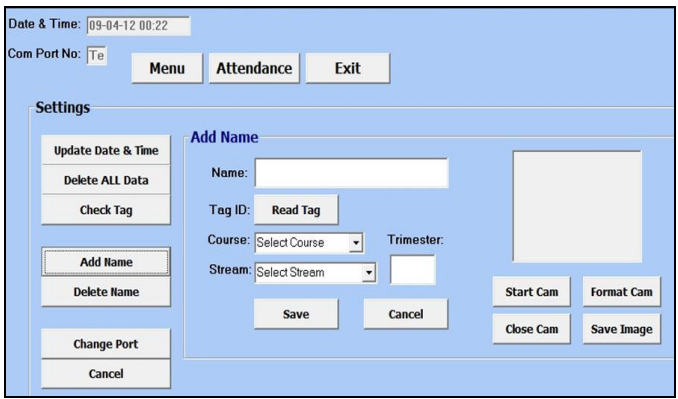

Fig. 5 GUI form to enter new student details

Figure 5 shows the new student registration page drafted using Visual Basic 6.0. The added functionality of capturing an image of the student provides visual authenticity whilst recording attendance. Data once stored in the database can only be modified by the system admin.

The RFID reader used in this research operates at a frequency of 125 KHz with an effective read range of 10cm only [13]. A short read range is preferred so as to maintain the authenticity and security of the attendance being recorded. Figure 6 depicts the display on the GUI as the system in the process of recording attendance. Data being recorded can be easily exported to a Microsoft Excel file for report generation. The database can be easily scaled to incorporate more details about the student.

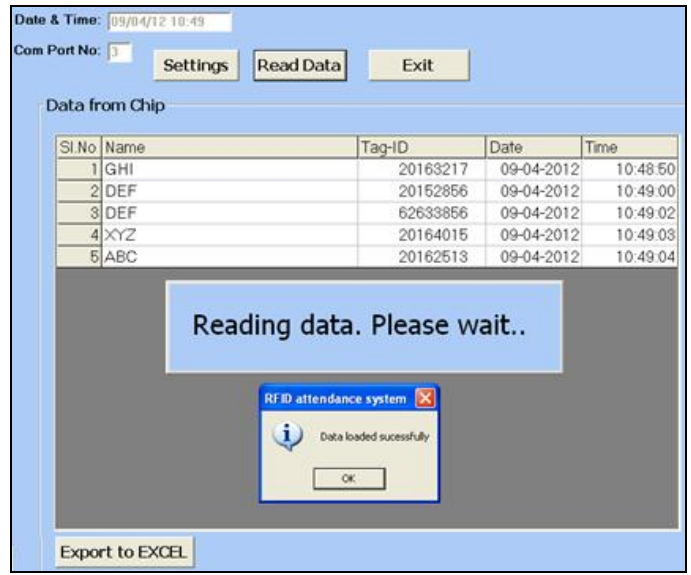

Fig 6 Recording student attendance

The overall system design is holistically depicted in figure 7, which is a block representation of the system. The figure shows the interconnection of two modules which are RFID module and Visual Basic 6.0 module. On the contrary, figure 8 displays the actual experimental setup of the research along with its individual components. The implementation of RFID technology in the system must be evaluated in a holistic to quantify its success.

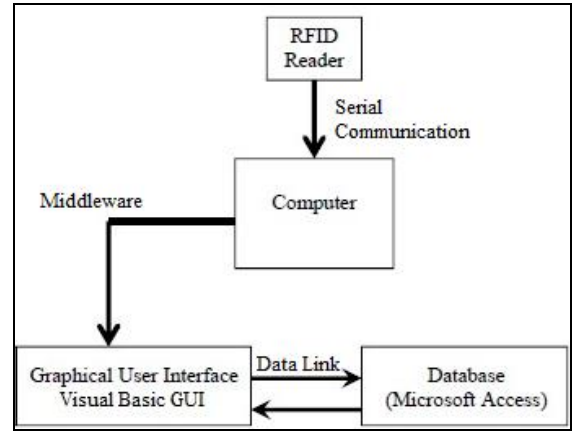

Fig. 6 Block representation of RFID system

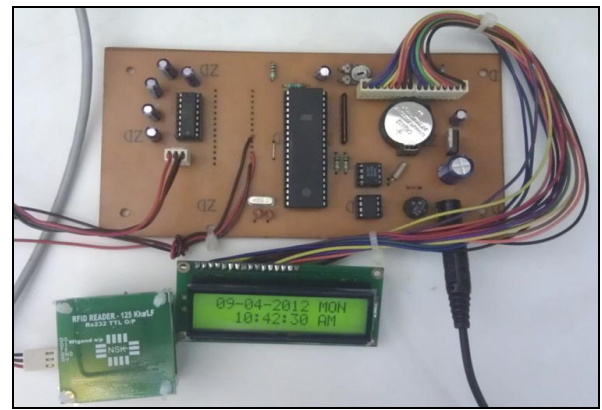

Fig. 7 Experimental setup

## 6. Results

The research was conducted on a sample of 60 students, enrolled in a particular course. The implementation of RFID technology has definitely quickened the entire of process of recording attendance. The traditional method of recording attendance involves individual manual entry; an arduous and a time consuming process. On average, based on experiment, the total time taken to record the attendance of a class of 60 students by manual entry method took approximately 10 minutes. This implies that approximately 10 seconds per student was required to record their attendance. This time duration includes visual and written authentication, after which the teacher records the attendance. In comparison (see figure 8), the total time taken for recording the attendance of 60 students using barcode and RFID technology is 120 seconds and 12 seconds respectively (see table 2). Based on the relationship obtained, a projection for a batch of 100 students is also forecasted.

## 7. Conclusion

The study has identified and explained the key benefits of RFID technology. RFID will open doors to a pool of applications from a plethora of industries [8]. Although the focal challenge to thwart the adoption is its investment cost, RFID technology provides an ocean of lucrative business opportunities that could convince several firms adopt it [14]. The first part of the paper explains the evolution of RFID technology and the role of its individual components within the system. The second part of the paper discusses the feasibility of employing RFID technology and how it is benefactor of improved efficiency at lowered costs. The last part of the paper highlights one of the numerous practical implementations of RFID technology.

RFID technology definitely promises an increased effectiveness and improved efficiency for business processes [22]. In the long run, with reducing unit tag and reader costs, several businesses will be able to leverage the benefits of RFID technology.

Table 2: Results of the Study

| Method | Total Number of Students | | | |
|---|---|---|---|---|
| | 1 | 10 | 60 | 100 |
| Manual Entry | 10 seconds | 100 seconds | 600 seconds | 1000 seconds |
| Bar Code | 2 seconds | 20 seconds | 120 seconds | 200 seconds |
| RFID technology | 0.2 seconds | 2 seconds | 12 seconds | 20 seconds |

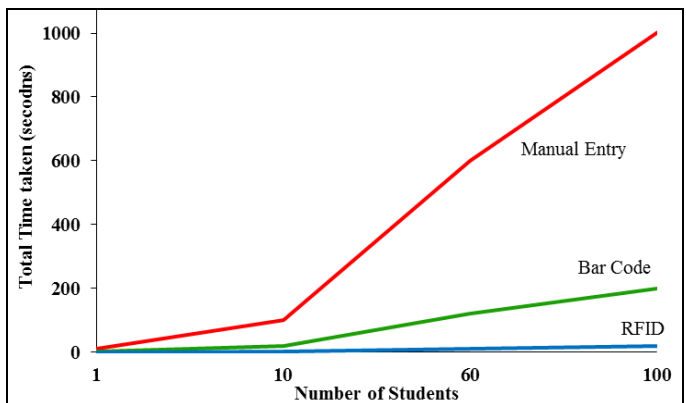

Fig 8 A line graph showing the comparison of total time taken to record the attendance of students.

As shown in table 2, compared with the time consumption in data entry for different technologies, RFID technology saves considerable amount of time and greatly improves the operation efficiency. Also with the adoption of this technology the process and product quality can be improved due to reduction in entry errors by manual human operations. Therefore, labour cost is reduced to perform the value added functions.